\documentclass[aps,prd,preprint]{revtex4}
\usepackage{epsfig}
\begin{document}
\draft
\title{On Hierarchy, Charge Universality, and 4D Effective Theory in Randall-Sundrum Models}

\author{
Katherine Benson\footnote{Electronic address:
benson@physics.emory.edu}}
\affiliation{Department of Physics, Emory University,
Atlanta, Georgia 30322-2430, USA}

\date{\today}

\begin{abstract}

We present a variant formulation of the Randall-Sundrum model which
solves both the hierarchy and charge universality problems.  We first
critique the rationale for hierarchy solution and 4D effective
interactions in the Randall-Sundrum model.  We note its asymmetric
treatment of matter and gravity in the warped braneworld background,
leaving uncalibrated the particle scale; as well as its unconventional
spatial attribution of integrated 4D effective gravity.  Matter and
massless gravitons {\em both} localize when branes form to warp
spacetime; thus consistent accounting of induced 4D physics must track
{\em both} particle and Planck scales through brane formation. We
perform such self-consistent tracking in the warped Randall-Sundrum
background, by treating matter as intrinsically extradimensional, on
par with gravity, with a unified mass scale. We find this definite,
self-consistent theory solves {\em two} major problems: the effective
4D theory shows robust hierarchy solution, and preserves charge
universality. Our unified 5D field theory lies at the Planck scale; it
induces an integrated 4D effective field theory with universal
charges, Planck scale 4D gravity, and {\rm TeV} scale matter. However,
this effective field theory describes, not 4D physics on a specific
brane, but 4D physics induced by an unobservably small ($12\ M_{\rm
Pl}^{-1}$) warped extra dimension. This unified approach validates
Randall-Sundrum hierarchy solution, while exemplifying a field theory
whose dimensional reduction preserves charge universality.

\end{abstract}

\vspace{0.5in}

\maketitle

\section{Introduction and Rationale\label{intro}}

Endowing a universe with spacetime dimensionality greater than 4 has
long enticed theorists, especially those seeking to reconcile gravity
and particle interactions. In the 1920s, Kaluza and Klein tried to
model electromagnetism as gravitational dynamics of a compactified
extra dimension.\cite{klein, kaluza} From the 1970's forward,
supergravity, string and M theories --- attempts to unify quantum
gravity and particle interactions --- attained consistency {\em only}
in higher dimension. Practitioners viewed the required extra
dimensions as compactified and small, negligibly affecting known
low-energy physics. In 1998, however, viability of {\em large}
compactified extra dimensions was noted. \cite{largeextra} Again extra
dimensions acted to reconcile gravity and particle interactions:
specifically, large extradimensional volume solved the hierarchy
problem, inducing a hierarchically large 4D Planck mass --- or
gravitational mass scale --- from extradimensional gravity at a
unified particle scale. Soon followed the viability of unbounded extra
dimension, when spacetime curves or ``warps'' around a ``3-brane'' ---
a four-dimensional subuniverse confining all standard
matter. \cite{rs2} Such branes, defects that confine standard matter,
while gravity explores a higher-dimensional ``bulk,'' are predicted by
string and M theory. \cite{polchinski, duff} In 5-dimensional toy
models, Randall and Sundrum showed first, that a brane's warping of
the extra dimension can bind massless gravitons to it, inducing an
effectively four-dimensional gravity coinciding observationally with
our own.  \cite{rs2} Second, they found the warped metric can induce a
hierarchical divergence between the particle scale and the effective
4D Planck scale on a brane.\cite{rs1} Thus extra dimension --- here
warped by the matter content of the 3-brane --- again acts to unify
gravity and particle interactions, inducing their observed hierarchy
from a unified extradimensional mass scale.

Randall-Sundrum's solution of the hierarchy problem spurred phenomenal
research activity.  This activity examined the effect of warped extra
dimension on particle phenomenology, on AdS-CFT issues within string
theory, and on cosmology. A copious literature establishes the
potential, in slightly generalized models, to resolve other
longstanding naturalness issues in cosmology and particle
phenomenology.

However, Randall-Sundrum's proposed hierarchy solution, while
suggestive, was incomplete. Their model began with the Horava-Witten
vacuum favored by M theory: with one large orbifolded extra dimension,
negative bulk cosmological constant, and 3-branes sitting at opposite
orbifold fixed points. For string theoretic reasons, ordinary
particles are trapped on the branes, where strings end. Gravity, and
in simple extensions, exotic particles, explore the full spacetime,
the bulk. To model this string-motivated picture, Randall-Sundrum
idealized the branes as infinitely thin, tension-carrying walls, with
ordinary matter intrinsically 4-dimensional and located on each
brane. Gravity they modeled, instead, as intrinsically
5-dimensional. They then showed that their positive-tension ``hidden''
brane confines massless gravitons, inducing from the original 5D
gravity a 4D effective field theory for gravity on the brane. This 4D
effective gravity, interpreted on the graviton-confining brane, has a
4D effective Planck scale comparable to its parent 5D Planck
scale. However, the view from the negative tension ``visible'' brane
differs. Here the 4D metric is exponentially suppressed, due to the
branes' extradimensional warping of the metric. Thus, when ``visible''
brane observers relate the induced 4D effective gravity to their local
4D metric, they perceive a hierarchically enhanced 4D Planck
scale. That is, the hierarchy problem is solved --- on the visible
brane only --- by a warp localizing massless gravitons far away,
so that an {\em integrated} 4D effective theory for gravity looks {\em
locally} like a hierarchically large 4D effective Planck scale.

This reasoning has gaps. First, it assigns to all of its fundamental
players, whether 5-dimensional gravity in bulk or 4-dimensional matter
on either brane, a single scale, a string scale set at our observed
{\rm TeV} particle scale. \cite{fn1} Yet the branes' existence
localizes the massless graviton as inexorably as it localizes matter;
and if this localization hierarchically shifts an effective scale for
localized gravitons, we must question whether the scale for 4D matter
particles, confined to either brane, itself flows trivially from the
string scale in a brane-warped geometry.  For, in taking as its
fundamental objects 4D particles and 5D gravity, the Randall-Sundrum
model takes objects which are, in fact, not equally fundamental. The
gravitons appear at some post-brane formation, but extradimensional
and pre-confined stage; while the particles' binding to the 3-brane is
assumed complete, cloaked behind the veil of string theory. Yet brane
formation inexorably binds both, at once: there is no evolutionary
stage where matter can be viewed as bound while graviton zero modes
are not. To examine hierarchy solution, a hierarchical divergence of
gravity and particle scales, we must track both scales from an
equivalent starting point: here, before {\em either} binds to
branes. Otherwise, our perceived hierarchy solution may arise from
partial accounting, from tracing shifts of the Planck scale while
ignoring those of the particle scale, in that single process when
branes form and 4D effective field theories on them emerge.

A second subtlety lies in the Randall-Sundrum interpretation of 4D
effective gravity. This effective theory is an integrated theory --- a
5D Planck action, integrated over the fifth dimension in the warped
brane background, to form a 4D effective Planck action.  Support for
the graviton zero mode coincides identically with support for this
extradimensional Planck action. With the graviton zero mode confined
to the hidden brane, the extradimensional Planck action integral also
receives support only on the hidden brane. One would consistently
regard this 4D effective gravity as salient only on the hidden brane,
much as we idealize an integrated brane tension as nonzero only
pointwise, at the location of a thin brane. Yet Randall-Sundrum treat
this 4D effective Planck action as if observers on branes fixed at any
point in the extra dimension all experience it identically; as if each
sees this same integrated 4D effective Planck action, then transcribes
it in terms of his own local metric, causing the perceived 4D Planck
mass to vary with position. This attribution of an integrated action
fully and simultaneously to all distinguishable points, without regard
to the action's support at each, is murky.

A third gap arises only from the spectacular success of the
Randall-Sundrum models. As noted, far-reaching and promising
consequences of such models were proclaimed, resolving longstanding
phenomenological and cosmological issues.  These results emerged from
the Randall-Sundrum toy models, with simple extensions (bulk scalar
fields --- themselves necessary to stabilize Randall-Sundrum I
\cite{Goldberger, Goldberger2}; bulk gauge fields \cite{Pomarol,
Carena}; adding cosmological energy density to Randall-Sundrum
backgrounds \cite{Langlois}; and detuning Randall-Sundrum backgrounds,
to allow nonvanishing effective 4D cosmological
constants. \cite{Langlois}). Like the original Randall-Sundrum models,
these models start at an ad hoc intermediate point, with ordinary 4D
matter already bound to the brane, with assumed ``standard''
interactions.  Only gravity and exotic particles explore the ``bulk'',
inducing effective 4D interactions on branes, and effective 4D
Friedmann equations, which are calculated. Consequences for brane
cosmology and phenomenology both stem from these induced 4D gravity
and exotic interactions. But, as argued for the hierarchy problem
above, these induced interactions or dynamics are intelligible only
when calibrated against ordinary particle interactions on the brane:
which may themselves arise or evolve nontrivially, when embedded in a
warped extradimensional geometry.

To address these issues, we must track {\em both} particle and gravity
scales consistently, back to a single unified scale predating either's
dynamical confinement onto 3-branes. We achieve this by modest
reinterpretation of the basic Randall-Sundrum scenario. We retain the
motivating context, the Horava-Witten vacuum with one large orbifolded
extra dimension, negative bulk cosmological constant, and 3-branes
sitting at opposite orbifold fixed points; we retain even the exact
Randall-Sundrum solution for the warped metric in this setting,
including its relationship of brane tensions $\tau_{\rm hid}$ and
$\tau_{\rm vis}$ and negative bulk cosmological constant $\Lambda_5$:
$\tau_{\rm hid} = -\tau_{\rm vis} = - \Lambda_5$. What we alter is the
relationship of matter and gravity to this warped geometry; we begin,
not at an asymmetric point of confined 4D matter and extradimensional
5D gravity, but at a symmetric point where {\em both} matter and
gravity are 5-dimensional, defined by conventional 5-dimensional
actions and field theories, with unified mass scale. The branes we
view as idealized defects, interacting only gravitationally through
their tension, which warps the metric, creating the Randall-Sundrum
warped background for unified 5D matter and gravity.

We take this extradimensional field theoretic approach to matter, not
because string theory {\em must} reduce to an effective field theory
for matter when it is still extradimensional, before 3-branes form,
although it could. We do this, instead, because the proposed hierarchy
solution, and the proposed phenomenological and cosmological
consequences, depend intricately on that transition from string theory
to effective field theory, which ends with 3-branes forming and
localizing {\em both} matter and gravity. Randall-Sundrum lifted one side of
a veil in that process, showing how a brane can bind gravitons and
shift effective Planck scales; yet the other side remains hidden,
obscuring the emergence of 4D matter and its interaction
scale. Assuming an extradimensional field theory for matter, exactly
on par with extradimensional gravity, provides a consistent and
calculable tool to lift the remaining veil, to extend the history of
the particle scale backward through its dynamical confinement onto 
branes. Whether field theory emerges as an effective theory for string
theory before 3-brane formation or not, it provides the only definite
and complete theoretical framework to ask our question, to
simultaneously peel away the veil on both sides. It provides a
rigorous, self-consistent testing ground, to see if the suggested
hierarchy mechanism might survive a full accounting, in a theory that
encodes {\em all} consequences of confinement in a brane-warped
geometry.

Moreover, extradimensional field theory shows promise as a tool for
implementing stringy predictions of brane formation and confinement.
Viewed solely as gravitational sources, branes can confine
extradimensional matter.  Their confinement of bulk matter is
inseparable from confinement of the graviton zero mode; both flow
inevitably from the brane's warping of the extradimensional
metric. Such gravitational confinement of extradimensional matter onto
4D domain walls has been explored before
\cite{Rubakov1,Rubakov2,Shifman}; generically, however, significant
challenges arise: first, in confining gauge fields; and second, in
confining matter while preserving 4D effective charge
universality. \cite{Rubakov}

In our interpretation, of extradimensional matter in a warped
Randall-Sundrum background, the self-gravitating branes accomplish
only some of the required stringy confinement. We find most matter
becomes confined when the branes form, although not all on the same
brane. Massive matter and gauge fields become confined on the visible
brane; massless matter fields, like massless gravitons, localize on
the hidden brane; and massless gauge fields remain delocalized, evenly
smeared throughout the bulk. This does not implement the anticipated
stringy vacuum, of complete gauged particle theories confined to
single branes. It does, however, define a sensible 4D effective field
theory, one which generates hierarchy even as it preserves charge
universality. This 4D effective field theory is intelligible {\em not}
as one localized on either brane; but instead as an integrated
effective field theory, over an extra dimension whose size ($12\
m_{\rm Pl}^{-1}$) remains too small to empirically resolve. In this
context, the integrated 4D effective field theory is the effective
theory perceived below near-Planckian energies. Only with
near-Planckian probes does the extra dimension resolve itself,
revealing distinguishable branes with distinct 4D effective theories
on each.

While this falls short of old domain wall localization goals, of
colocalizing an entire standard particle theory while preserving
charge universality, its achievements are substantial. It accomplishes
the elusive goal of preserving charge universality in an integrated
effective field theory. Simultaneously, it generates the known
hierarchy between particle and Planck scales. These accomplishments
arise rigorously, entirely within a calculable field theory. They show
that the Randall-Sundrum mechanism for generating hierachy via warped
extra dimension can be robust, emerging from the self-consistent flow
of all mass scales during brane formation; and that warped extra
dimension has sufficient structure to force universality in effective
charge integrals. The resulting 4D effective field theory is exactly
what we want; it's just not what we expected. From the Horava-Witten
heuristic, we expected the relevant 4D effective field theory to be
local, confined to a single brane in a large extra dimension; instead
we find it to be integrated, describing smeared effects of an entire,
but unobservably small, warped extra dimension.

We describe the details of this dimensional reduction as follows. We
define our unified 5D theory in the warped Randall-Sundrum background
in section \ref{details}. We then examine the Kaluza-Klein wave
functions and spectra determined by this warped background, and the
resulting integrated 4D effective field theories, for gravity (section
\ref{det:grav}), free scalars and fermions (section \ref{det:higgs}),
and free gauge fields (section \ref{det:gauge}). These sections
establish the hierarchy, with 4D effective particle masses undergoing
hierarchical suppression. Finally, we examine 4D effective interactions,
particularly the emergence of effective charge universality, in section
\ref{det:interp}. We conclude  in section \ref{conclude}.

\section{Details\label{details}}

We begin with the warped Randall-Sundrum background, whose orbifolded
extra dimension has hidden and visible 3-branes at fixed points $\phi
= 0$ and $\phi = \pi$ respectively, with tensions related to the
negative bulk cosmological constant $\Lambda_5$ by $\tau_{\rm hid} =
-\tau_{\rm vis} = - \Lambda_5$. The stress-energy of this background
determines a warped metric solution to Einstein's equations,
\begin{equation}
ds^2      = e^{-2kR|\phi|}\ \bar{g}_{\mu\nu}dx^\mu dx^\nu  +R^2d\phi^2\ \ .
\label{eq=metric}
\end{equation}
Here Greek indices $\mu\nu$ index the 4 Lorentz directions; $k$ is of
order $M_{\rm Pl}$; $R$ is the compactification radius for the extra
dimension; $\phi$ ranges from $-\pi$ to $\pi$, with $\phi$ and $-\phi$
identified by orbifolding; and we take the mainly $+$ sign
convention. For the Randall-Sundrum solution, $\bar{g}_{\mu\nu} =
\eta_{\mu\nu}$; we retain the broader notation here to clarify the
induced 4D Planck action.

We consider field theory in this warped background, where both
gravity and matter are intrinsically 5-dimensional, at a unified mass
scale. To establish a 5D particle scale, we assume that electroweak
symmetry-breaking occurs in the bulk.  Note that even in the
brane-warped background, homogeneous electroweak symmetry-breaking
throughout the bulk remains a solution; and this simplest solution,
where the 5D Higgs field assumes a constant vev, is the one we
adopt. Thus in our paradigm electroweak symmetry is broken in the
bulk, where the Higgs field $H$ lives as a 5D massive scalar doublet,
and $W^{\pm}$ and $Z$ as 5D massive gauge bosons. Their 5D
Higgs-induced masses coincide with the 5D Planck scale.  How these
particle masses shift, when 4D effective field theory emerges in the
brane-warped background, sets the flow of the particle scale during
brane formation. More specifically, the lightest Kaluza-Klein modes
induced for $H$, $W^{\pm}$ and $Z$ become the apparent 4D Higgs and
weak gauge bosons, setting the induced particle scale.  The effective
4D Planck scale follows instead from  normalization of the graviton
zero mode, as calculated in \cite{rs2}. Note that in its basic
structure, this warped dimensional reduction provides a blueprint for
hierarchy solution, since the flow of Planck and particle scales from 5
to 4 dimensions obey distinct mechanisms.  These unrelated mechanisms
--- graviton zero mode normalization, versus particle lowest mass
eigenvalue --- may lead to divergent flows, inducing from a unified
bulk scale hierarchically divergent Planck and particle scales. This,
we argue below, is what occurs under dimensional reduction in the
warped Randall-Sundrum background (\ref{eq=metric}).

Concretely, we take as our action for gravity and matter
\begin{equation}
{\cal S} = \int d^4x \ \int_{-\pi}^{\pi}\ d\phi\ \sqrt{-g}\ \left(\ 
\begin{array} {c}
(2M_{\rm Pl, 5}^3\, {\cal R}-\Lambda_5)\  -\ 
(g^{AB}\partial_A H^\dagger \partial_B H + m_{5}^2\, H^\dagger H)\\[5pt] 
- \sum_a\ (\ \frac{1}{4}\ g^{AC}g^{AD}F_{a\, AB}F_{a\, CD} + \frac{m_{a,5}^2}{2}\ g^{AB}A_{a\, A}A_{a\, A})\ +\ {\cal L}_\psi\ +\ {\cal L}_{\rm int}
\end{array}  \right)\,.
\label{eq=action}
\end{equation}
Here Roman indices $A,\ B,\ C,\ D$
range over all five dimensions $x^\mu, \phi$; $g_{AB}$ is the warped
metric (\ref{eq=metric}) and ${\cal R}$ its 5D Ricci scalar; and
$M_{\rm Pl, 5}$ and $\Lambda_5$ are the 5D Planck mass and
cosmological constant. $H$ is the 5D Higgs doublet and $A_{a\, A}$ are
the 5D gauge fields corresponding to the massless photon and massive
weak gauge bosons. The weak scale masses $m_{5}$ and $m_{a,5}$
for $Z$ and $W^{\pm}$ lie at the Planck scale $M_{\rm Pl, 5}$, set at
our own Planck scale $10^{19}\ {\rm GeV}$. ${\cal L}_\psi$ is the
usual Lagrangian for fermions with broken electroweak symmetry; and
${\cal L}_{\rm int}$ the usual electroweak interaction Lagrangian
(including charge interaction terms).

We now find Kaluza-Klein modes in the background (\ref{eq=metric}),
and the resulting integrated 4D effective field theories, for gravity;
for free scalar and Fermi fields; and for free gauge bosons in turn. All have
been examined by previous authors, in differing contexts, whose
results we review briefly. In section \ref{det:interp}, we consider
charged interactions in the 4D effective field theory, and show how
charge universality arises. Finally, we offer a self-consistent
interpretation of the full integrated 4D effective field theory ---
for gravity as well as matter, with hierarchy and charge universality
--- as a 4D effective theory due to unobservably small warped extra
dimension.

\subsection{Gravity\label{det:grav}}

Randall-Sundrum's original papers \cite{rs2,rs1} themselves analyzed
the 4D gravity induced by the gravitational action (\ref{eq=action})
in the RS background (\ref{eq=metric}); the analysis of
\cite{Goldberger}, in the massless limit, applies as well. We note
relevant results, using a distinct approach to highlight the relation
of the induced 4D Planck scale to the graviton zero mode's
normalization.

Gravitons, as identified by a 4D observer, correspond to deformations
of the metric (\ref{eq=metric}) by $ \bar{g}_{\mu\nu}\rightarrow
\bar{g}_{\mu\nu} + \bar{h}_{\mu\nu}$.  We take for $\bar{h}_{\mu\nu}$
a Kaluza-Klein decomposition of modes,
\begin{equation}\bar{h}_{\mu\nu} = \sum_n\ \psi_n(x^\mu)\ \bar{h}_n(\phi)\
\hat{e}_{\mu\nu}\,,\end{equation} 
with $\psi_n(x^\mu)$ a 4D plane wave of mass $m_n$. Einstein's
equations, which induce an equation of motion for $\bar{h}_{\mu\nu}$,
induce for the Kaluza-Klein graviton mode $\bar{h}_n(\phi)$,
\begin{equation}
\frac{1}{R}\ e^{2kR|\phi|}\ \frac{d}{d\phi}\
\left( \ R^{-1}\  e^{-4kR|\phi|}\ \frac{d\bar{h}_n}{d\phi} \right) \ =\ 
- m_n^2\, \bar{h}_n\, .
\label{eq=gravSL}
\end{equation}
This is a Sturm-Liouville equation for $\bar{h}_n$, with weight
$\rho = R\, e^{-2kR|\phi|}$ and lowest eigenvalue $m_n = 0$. This graviton zero
mode has one regular solution: the constant solution $\bar{h}_o =
1/\sqrt{N}$. The normalization factor $N$ is given by
\begin{equation}
N = \int_{-\pi}^{\pi} \ \rho d\phi\ =\ R\,\int_{-\pi}^{\pi} \ e^{-2kR|\phi|}\ d\phi = \frac{1}{k}\ (\ 1- e^{-2kR\pi}\ )\ .
\label{eq=zmnorm}
\end{equation}
Note that the zero mode's spatial probability distribution is given by
$\rho |\bar{h}_o|^2$, falling as $e^{-2kR|\phi|}$. Thus we interpret
the graviton zero mode as confined to the hidden brane at $\phi = 0$.

The normalization factor $N$ governs flow of the Planck scale, from
its bulk value to its effective 4D value on the hidden brane. We see this by
calculating the effective 4D Planck action induced by the action
(\ref{eq=action}) in the warped background metric (\ref{eq=metric}).
Key is the fact that, for unconstrained 4D metric $\bar{g}_{\mu\nu}$,
the 5D Ricci scalar ${\cal R}$ depends simply on the 4D Ricci scalar
$\bar{\cal R}$ induced by $\bar{g}_{\mu\nu}$:
\begin{equation}
{\cal R} = e^{2kR|\phi|}\ \bar{\cal R}\ \ + \ \ \mbox{{\rm $\bar{g}_{\mu\nu}$-independent terms}} \ ,
\end{equation}
while the 5D measure $\sqrt{-g} = R\,e^{-4kR|\phi|}\ \sqrt{-{\bar{g}}}$.
Thus the 5D Planck action in (\ref{eq=action}) determines an effective 4D
Planck action
\begin{equation}
{\cal S}_{g,4} \ =\ \int d^4x \ \sqrt{-\bar{g}}\ \ 
2M_{\rm Pl, 5}^3\ \bar{\cal R}\ \int_{-\pi}^{\pi}\ Rd\phi\  e^{-2kR|\phi|} \ =\ 
N\ 2M_{\rm Pl, 5}^3\ \int d^4x \ \sqrt{-\bar{g}}\ 
 \bar{\cal R}\ \ .
\end{equation}
This corresponds to a 4D Planck action defined solely in terms of the 4D
metric $\bar{g}_{\mu\nu}$, with effective 4D Planck mass
\begin{equation}
M_{\rm Pl, 4}^2 = N\ M_{\rm Pl, 5}^3 = \frac{1}{k}\ 
(\ 1- e^{-2kR\pi}\ )\ M_{\rm Pl, 5}^3 \,,
\label{eq=Planckm}
\end{equation}
as noted in \cite{rs2}. As promised in Section \ref{intro}, the 5D
Planck action has extradimensional support identical to that of the
graviton zero mode. Thus the integrated 4D Planck action is
consistently viewed as either (1) being localized on the hidden brane
at $\phi = 0$, or (2) attributable diffusely to the entire bulk, {\em
only} if position in the extra dimension is unresolvable.

Ultimately we choose $kR \approx 12$, as in \cite{rs1}. This means
that the 4D effective Planck mass $M_{\rm Pl, 4}$ is little changed
from its parent value $M_{\rm Pl, 5}$. This could describe 4D gravity
on the hidden brane, for a large extra dimension; or 4D gravity over
the full space, for unresolvably small extra dimension. Here, since
$R \approx 12\ M_{\rm Pl}^{-1}$, we fall into the second category, of
4D effective gravity due to an unresolvable extra dimension.

The massive Kaluza Klein gravitons have a discrete spectrum. 
Their eigenfunctions
\begin{equation}
\bar{h}_n \sim e^{2kR|\phi|} \ \left[\ J_2\left(\ \frac{m_n}{k}\ e^{kR|\phi|}
\right) + \frac{\pi m_n^2}{4k^2}\ Y_2\left(\ \frac{m_n}{k}\ e^{kR|\phi|}\right)\ \right]
\label{eq=gravKKs}
\end{equation}
have Kaluza-Klein masses
\begin{equation}
{m}_n = kx_n\, e^{-kR\pi}
\label{eq=gravmn}
\end{equation}
where $x_n$ obey the condition $2J_2(x_n) + x_n\, J_2'(x_n) = 0$, with
$x_n$  of order 1 for lowlying modes. The exponential suppression
yields Kaluza-Klein gravitons with {\rm TeV}-scale masses.
 
The lowlying modes are dominated by the $J_2$ solution, so that their
probability distribution $\rho |\bar{h}_n|^2$ grows as $e^{2kR|\phi|}
\ J_2^2(\ x_n\ e^{kR(|\phi|-\pi)})$. Thus massive graviton
Kaluza-Klein modes localize on  the visible brane at $\phi = \pi$.

\subsection{Higgs and Fermions\label{det:higgs}}

Bulk scalar fields in the brane-warped background (\ref{eq=metric})
were first analyzed by \cite{Goldberger}, whose results we reprise
here.  We take for the 5D scalar Higgs $H$ in action (\ref{eq=action}) 
a Kaluza-Klein decomposition of modes,
\begin{equation}H = \sum_n\ \psi_n(x^\mu)\ H_n(\phi)\,,
\label{eq=HKKdecomp}\end{equation}
with $\psi_n(x^\mu)$ again a 4D plane wave of mass $m_n$. The field equations for $H$ yield
\begin{equation}
\frac{1}{R}\ e^{2kR|\phi|}\ \frac{d}{d\phi}\
\left( \  R^{-1}\ e^{-4kR|\phi|}\ \frac{dH_n}{d\phi} \right) \ -\  m_{5}^2\ e^{-2kR|\phi|}\  H_n\ =\ 
- m_n^2\, H_n\, ,
\label{eq=HiggsSL}
\end{equation}
again a Sturm-Liouville equation with weight $\rho = R\,
e^{-2kR|\phi|}$. For massless scalars, this coincides with the
graviton Kaluza-Klein decomposition, giving identical spectra: a
massless scalar localized on the hidden brane, and excited ${\rm
TeV}$-scale Kaluza Klein modes localized on the visible brane. For our
massive 5D Higgs, however, there are only massive solutions
\begin{equation}
H_n \sim e^{2kR|\phi|} \ \left[\ J_\nu\left(\ \frac{m_n}{k}\ e^{kR|\phi|}
\right) + b_{n\nu}\ Y_\nu\left(\ \frac{m_n}{k}\ e^{kR|\phi|}\right)\ \right]
\label{eq=HiggsKKs}
\end{equation}
where $\nu = \sqrt{4 + m_{5}^2/k^2}$. These
have Kaluza-Klein masses
\begin{equation}
{m}_n = kx_{n\nu}\, e^{-kR\pi}
\label{eq=Higgsmn}
\end{equation}
where $x_{n\nu}$ obey the condition 
\begin{equation}\label{eq=Higgsxcrit}2J_\nu(x_{n\nu}) + x_{n\nu}\,J_\nu'(x_{n\nu}) = 0\ ,\end{equation} with
$x_{n\nu}$  of order 1 for lowlying modes. The exponential suppression
yields a Kaluza-Klein spectrum with {\rm TeV}-scale masses.
Lowlying modes are dominated by the $J_\nu$ solution, so that their
probability distribution $\rho |H_n|^2$ grows as $e^{2kR|\phi|}
\ J_\nu^2(\ x_{n\nu}\ e^{kR(|\phi|-\pi)})$. Thus massive scalar Higgs
Kaluza-Klein modes all localize on  the visible brane at $\phi = \pi$, with 
{\rm TeV}-scale masses.

The lowest Higgs Kaluza-Klein mode, which we perceive as the 4D Higgs,
thus has a {\rm TeV} scale mass. In a resolvable extra dimension, this
4D Higgs would live only on the visible brane; but here with $R
\approx 12\ M_{\rm Pl}^{-1}$, we perceive the scalar Kaluza-Klein 4D
effective theory as a holistic, non-localized consequence of
unresolved extra dimension.

Note that massive fermions, despite additional structure, obey the
same Klein-Gordon field equations as the massive scalars. Thus their
Kaluza-Klein spectrum and localization are also determined by these
results; their binding to the visible brane is driven by gravity, not
by direct couplings to condensed fields. 

Distinct scalar and fermion species differ in Kaluza-Klein structure
{\em only} due to differing bulk mass. Consequences of varying nonzero
bulk mass are muted, however, shifting only the order $\nu$ of the
Bessel function solutions, rather weakly for nonhierarchically related
differing masses. This alters the mass spectrum weakly, through the
exact placement of zeroes $x_{n\nu}$; it also alters, in detail, the
spatial profile $H_n(\phi)$. However, as we show in section
\ref{det:interp}, the varied spatial profiles of distinct species
do not alter the 4D induced effective charge.

\subsection{Gauge Fields\label{det:gauge}}

Gauge fields in the brane-warped background (\ref{eq=metric}) were
first analyzed by \cite{Pomarol}, whose results we review here.  We
consider the 5D massive gauge field $A_{a\, A}$ in action
(\ref{eq=action}). We make the gauge choice $A_{a\, \phi} = 0$, exhausting 
extradimensional gauge freedom only, and define a 4D effective gauge field $\bar{A}_{a\, \mu}$, 
as perceived by the 4D observer, by 
\begin{equation}\bar{A}_{a\, \mu} = A_{a\, \mu};\ 
\bar{A}^{a\, \mu} = \bar{g}^{\mu\nu}\, \bar{A}_{a\, \nu} = e^{2kR|\phi|}\ 
A^{a\, \mu}\ \ .
\label{eq=4DAdef}\end{equation}
We decompose $\bar{A}_{a\, \mu}$ as follows:
\begin{equation}\bar{A}_{a\, \mu} = \sum_n\ \bar{A}_{a\, \mu\,n}(x^\nu)\ {\cal A}_{an}(\phi)\,,
\label{eq=AKKdecomp}\end{equation}
where $\bar{A}_{a\, \mu\, n}(x^\nu)$ obeys a (linearized) 4D massive gauge
field equation,
\begin{equation}
\frac{1}{\sqrt{-\bar{g}}}\ \partial_\mu\,\left[\ \sqrt{-\bar{g}}\
\bar{g}^{\mu\rho}\,\bar{g}^{\nu\sigma}\, \bar{F}_{a\,\rho\sigma\, n}\
\right] = m_{a\, n}^2\,\bar{A}_{a\,n}^{\, \nu}\ \ .\end{equation} These choices, along with the choice $\rho = R$ in Equation (\ref{eq=gaugeSL}) below,  insure a 4D
gauge Kaluza-Klein decomposition: that is, the 5D gauge action
(\ref{eq=action}) induces, after $\phi$-integration, an effective
theory which is a sum over  4D actions for massive 4D gauge
fields $\bar{A}_{a\,n}^{\, \mu}$, in the metric $\bar{g}_{\mu\nu}$.

The (linearized) 5D massive gauge field equation for ${A}_{a}^{\, \mu}$
induces, for the modulating wave function ${\cal A}_{an}(\phi)$,
\begin{equation}
\frac{1}{R}\ \frac{d}{d\phi}\
\left( \ R^{-1} e^{-2kR|\phi|}\ \frac{d{\cal A}_{an}}{d\phi} \right) \ -\  m_{i,5}^2\ e^{-2kR|\phi|}\  {\cal A}_{an}\ =\ 
- m_{an}^2\, {\cal A}_{an}\, ,
\label{eq=gaugeSL}
\end{equation}
a Sturm-Liouville equation with weight $\rho = R$.

For massless gauge fields, we have lowest eigenvalue $m_{\gamma o} = 0$. This
photon zero mode, which we perceive as the 4D photon, has one regular
solution: the constant solution ${\cal A}_{\gamma o} = 1/\sqrt{2\pi R}$.  Note
that the photon's spatial distribution, given by $\rho |\bar{h}_o|^2$,
is here constant: the photon is smeared evenly throughout the bulk.

The massive gauge Kaluza Klein modes have
eigenfunctions
\begin{equation}
{\cal A}_{an} \sim e^{kR|\phi|} \ \left[\ J_{\nu '}\left(\ \frac{m_{an}}{k}\ e^{kR|\phi|}
\right) + b_{an{\nu '}}\ Y_{\nu '}\left(\ \frac{m_{an}}{k}\ e^{kR|\phi|}\right)\ \right]
\label{eq=gaugeKKs}
\end{equation}
where ${\nu '} = \sqrt{1 + m_{\rm a,5}^2/k^2}$. These
have Kaluza-Klein masses
\begin{equation}
{m}_{an} = kx_{an{\nu '}}\, e^{-kR\pi}
\label{eq=gaugemn}
\end{equation}
where $x_{an{\nu '}}$ obey the condition 
\begin{equation}\label{eq=gaugexcrit}J_{\nu '}(x_{an{\nu '}}) +
x_{an{\nu '}}\,J_{\nu '}'(x_{an{\nu '}}) = 0\ ,\end{equation}
with $x_{an{\nu '}}$ of
order 1 for lowlying modes. As for scalar fields, exponential
suppression yields a Kaluza-Klein gauge spectrum with {\rm TeV}-scale
masses, completing the claimed hierarchy generation between
effective Planck and particle scales.  Lowlying modes are dominated by
the $J_{\nu '}$ solution, so that their probability distributions
$\rho |{\cal A}_{an}|^2$ scale as $J_{\nu '}^2(\ x_{an{\nu '}}\
e^{kR(|\phi|-\pi)})$. Thus massive gauge Kaluza-Klein modes are also
localized mainly on the visible brane at $\phi = \pi$.

Note that we perceive the $n=1$ Kaluza-Klein modes as our
effective 4D weak gauge bosons.  These have {\rm TeV} scale masses,
and localize on the visible brane. Here again though, with $R \approx 12\
M_{\rm Pl}^{-1}$ unresolved, we perceive the Kaluza-Klein 4D
effective gauge theory as a holistic, non-localized consequence
of small warped extra dimension.

As for massive scalars and fermions, the massive weak gauge bosons
$W^\pm$ and $Z$ differ in Kaluza-Klein structure {\em only} due to
their differing bulk mass. This mildly shifts the order ${\nu '}$ of
the Bessel function solutions, causing distinct effective 4D masses,
by distinguishing the exact zeroes $x_{a1{\nu '}}$. Again varying bulk
mass alters in detail the spatial profile ${\cal A}_{an}(\phi)$; this does have some consequence for induced 4D effective charge couplings, as we
explore below.

\subsection{Effective Charges and Charge Universality\label{det:interp}}

The sections above detail how extradimensional integration of the 5D action
(\ref{eq=action}) yields normalized 4D effective field theories for
Kaluza-Klein spectra of free gravitons, scalars and fermions, and gauge
fields. What does extradimensional integration of these Kaluza-Klein
modes imply for their effective charge interactions?

Consider the  5D charge interaction for charged fermions:
\begin{equation}
{\cal S_{\rm int}} = \int d^4x \ \int_{-\pi}^{\pi}\ d\phi\ \sqrt{-g}\ 
e_{a5}\, \bar{\psi} \not{\!\!A}_aT_a\psi \ \ ,
\end{equation}
where $\psi$ and $A_a$ are intrinsically 5D fields. We use the
Kaluza-Klein decompositions (\ref{eq=HKKdecomp}) and (\ref{eq=AKKdecomp})
for $\psi$ and $A_a$ respectively, noting that our 4D gauge field
definition for $\bar{A}_a$, equation (\ref{eq=4DAdef}), implies $
\not{\!\!A}_a = e^{2kR|\phi|}\, \bar{\not{\!\!A}}$. Since the 5D measure $\sqrt{-g} = R\,e^{-4kR|\phi|}\ \sqrt{-{\bar{g}}}$, we may perform the extradimensional integration to obtain
\begin{equation}
{\cal S_{\rm int}} = \sum_{lmn}\ \int d^4x \ \sqrt{-\bar{g}}\ 
e_{a5}\, \bar{\psi}_l {\not{\!\!\bar{A}}_{am}}T_a\psi_n \ {\cal M}_{lmn}
\end{equation}
where 
\begin{equation}
{\cal M}_{lmn} \equiv \int_{-\pi}^{\pi}\ d\phi\ Re^{-2kR|\phi|}\ H_l^*{\cal A}_{am} H_n \ \ .
\label{eq=qeffdef}
\end{equation}

First, consider effective charge couplings to the 4D photon. Here
${\cal A}_{am}$ is the photon zero mode solution ${\cal A}_{\gamma o} = 1/\sqrt{2\pi R}$. Since the
photon's extradimensional profile is constant, 
\begin{equation}
{\cal M}_{l0n} = \frac{1}{\sqrt{2\pi R}}\  \int_{-\pi}^{\pi}\ d\phi\ Re^{-2kR|\phi|}\ H_l^* H_n \ = \ \frac{1}{\sqrt{2\pi R}}\ \delta_{ln},
\end{equation}
using the Sturm-Liouville normalization $Re^{-2kR|\phi|}$ for
$H_n$, from section {\ref{det:higgs}. Thus induced electromagnetic charge is universal, and related
to 5D electromagnetic charge by $e_{\rm 4D} = e_5/\sqrt{2\pi R}$.

For charged couplings to massive gauge bosons, we must examine
Kaluza-Klein eigensolutions $H_n$ (given by equation
(\ref{eq=HiggsKKs})) and ${\cal A}_{am}$ (given by equation
(\ref{eq=gaugeKKs})) more closely. We approximate each by their
$J_\nu$ (or $J_{\nu'}$) components, hierarchically dominant for
lowlying modes. Then
\begin{eqnarray}
{\cal A}_{am} &\approx& {\rm a}_{am}\, e^{kR|\phi|}\, J_{\nu'}(x_{am\nu'}e^{kR(|\phi|-\pi)}) \nonumber\\
H_n &\approx& {\rm h}_n\, e^{2kR|\phi|}\, J_{\nu}(x_{m\nu}e^{kR(|\phi|-\pi)})\ ,
\label{eq=approxmodes}
\end{eqnarray}
with $x_{am\nu'}$ and $x_{m\nu}$ order one zeroes determined by
equations (\ref{eq=Higgsxcrit}) and (\ref{eq=gaugexcrit}),
respectively. We normalize with the Sturm-Liouville weights $R$ for
${\cal A}_{am}$, and $Re^{-2kR|\phi|}$ for $H_n$, as discussed in
sections {\ref{det:higgs} and {\ref{det:gauge}. This gives
\begin{eqnarray}
{\rm a}_{am}^{-2} &=&  \int_{-\pi}^{\pi}\ d\phi\ Re^{2kR|\phi|}\ J_{\nu'}^2(x_{am\nu'}e^{kR(|\phi|-\pi)}) \nonumber\\
{\rm h}_n^{-2} &=&  \int_{-\pi}^{\pi}\ d\phi\ Re^{2kR|\phi|}\ J_{\nu}^2(x_{n\nu}e^{kR(|\phi|-\pi)}) \ \ .
\end{eqnarray}
Transforming to the variable $w = e^{kR(|\phi| - \pi)}$, and noting $e^{-kR\pi} \approx 0$, the integrals become
\begin{eqnarray}
{\rm a}_{am}^{-2}= \frac{2}{k} \ {e^{2kR\pi}}\ \int_0^1\ w J_{\nu'}^2(x_{am\nu'}w)\ dw\ \equiv \frac{2}{k} \ {e^{2kR\pi}}\ I_{\nu'}( x_{am\nu'})\nonumber\\[5pt]
{\rm h}_{n}^{-2}= \frac{2}{k} \ {e^{2kR\pi}}\ \int_0^1\ w J_{\nu}^2(x_{n\nu}w)\ dw\ \equiv \frac{2}{k} \ {e^{2kR\pi}}\ I_{\nu}( x_{n\nu})
\ .
\label{eq=norms}
\end{eqnarray}

The normalization integral $I_\alpha(a)$ can be done by noting, from
Bessel's equation, that 
\begin{equation}
\int_0^1\ w J_\alpha (aw) J_\alpha (bw) = \frac{1}{b^2 - a^2}\ \left(\ J_\alpha(b) aJ_\alpha '(a) - J_\alpha(a) bJ_\alpha '(b)\ \right)\ \ .
\end{equation}
For $a$ and $b$ distinct zeroes obeying 
\begin{equation}\label{eq=genxcrit}
x{J_\alpha '} (x) = - \beta J_\alpha (x)\ , \end{equation}
our criteria for fermions (with $\alpha = \nu,\ \beta = 2$)  and for gauge modes (with $\alpha = \nu',\ \beta = 1$), we of course get orthogonality. Taking the limit $b\rightarrow a$, using l'Hopital's rule, gives
\begin{equation}
\label{eq=defI}
I_\alpha(a) \equiv \int_0^1\ w J_\alpha^2 (aw) = \frac{1}{2}\ \left[\ {J_\alpha '}^2 (a)
+ \left(1 - \frac{\alpha^2}{a^2}\ \right)\, J_\alpha^2(a)\ \right]\ ,\end{equation}
using Bessel's equation only. For the criterion (\ref{eq=genxcrit}),
this gives the specific result
\begin{equation}
\label{eq=specI}
I_\alpha(a) = \frac{1}{2}\ {J_\alpha}^2 (a) \left(\ 1
- \frac{\alpha^2 - \beta^2}{a^2}\ \right)\ \ .\end{equation} Note in either case,  $\alpha^2 -
\beta^2 = m_5^2/k^2$, where $m_5$ is the 5D mass $m_{5}$ for the
fermion, $m_{a,5}$ for the gauge modes. This yields the normalization factors
\begin{eqnarray}
{\rm a}_{am}^{-2} &=& \frac{1}{k}\ e^{2kR\pi} \ {J_{\nu'}}^2 (x_{am\nu'})  
 \left(\ 1 - \frac{m_{a,5}^2}{k^2x_{am\nu'}^2}\ \right)
\nonumber\\[5pt]
{\rm h}_n^{-2} &=& \frac{1}{k}\ e^{2kR\pi} \ {J_{\nu}}^2 (x_{n\nu})  
 \left(\ 1 - \frac{m_5^2}{k^2x_{n\nu}^2}\ \right)  \ \ .
\end{eqnarray}
Note that this corrects the gauge mode normalization of Pomarol
\cite{Pomarol} by the factor in parentheses.

The normalization integral $I_\alpha(a)$ has a useful property,
\begin{equation} 
\left(\ 2 + a\ \frac{\partial\ }{\partial a}\ \right)\  I_\alpha(a)  = J_\alpha^2(a)
\label{eq=Iode}
\end{equation}
This follows from differentiating the integral form for $I_\alpha(a)$ with
respect to $a$, then integrating by parts, and can be verified for the
generic solution for $I_\alpha(a)$ in equation (\ref{eq=defI}).

We now proceed to the effective charge integral ${\cal M}_{nmn}$. For
our modes, from equations (\ref{eq=qeffdef}) and (\ref{eq=approxmodes}),
\begin{eqnarray}
{\cal M}_{nmn} &=& 2 {\rm a}_m {\rm h}_n^2 \int_0^{\pi}\ d\phi\ Re^{3kR\phi}\ 
J_{\nu'}(x_{am\nu'}e^{kR(\phi-\pi)})\ J_{\nu}^2(x_{n\nu}e^{kR(\phi-\pi)})\nonumber\\
&=& \frac{2}{k} \  {\rm a}_m {\rm h}_n^2\ e^{3kR\pi}\ \int_0^1\ dw\,w^2\, J_{\nu'}(x_{am\nu'}w) J_{\nu}^2(x_{n\nu}w) \ ,
\end{eqnarray}
after the change of variables $w = e^{kR(\phi - \pi)}$. 

We now consider the integral 
\begin{equation}
J(b, a) \equiv \int_0^1\ dw\,w^2\, J_{\nu'}(bw) J_{\nu}^2(aw) \ .\end{equation}
Integrating by parts,
\begin{eqnarray}
J(b, a) &=& \left. \frac{w^3}{3}\ J_{\nu'}(bw) J_{\nu}^2(aw)\ \right|_0^1 
- \frac{1}{3} \ \int_0^1 \ dw\,w^3\ \left(\ b J_{\nu'}'(bw) J_{\nu}^2(aw) + 
 2aJ_{\nu'}(bw) J_{\nu}(aw)J_{\nu}'(aw)\ \right) \nonumber\\[5pt]
&=& \frac{1}{3}\ J_{\nu'}(b) J_{\nu}^2(a)- \frac{1}{3} \ \left(\ a\ \frac{\partial\ }{\partial a} +b\ \frac{\partial\ }{\partial b}\ \right) \ J(b, a)\ \ .
\end{eqnarray}
Thus the integral $J(b,a)$ obeys the partial differential equation
\begin{equation}
\left(\ 1 + \frac{1}{3} \ \left(\ a\ \frac{\partial\ }{\partial a} +b\ \frac{\partial\ }{\partial b}\ \right)\ \right) \ J(b, a) =  \frac{1}{3}\ J_{\nu'}(b) J_{\nu}^2(a)\ \ ,
\end{equation}
or, more suggestively,
\begin{equation}
\left(\ \left(\ 2 + a\ \frac{\partial\ }{\partial a}\ \right) + \left(\ 1 + b\ \frac{\partial\ }{\partial b}\ \right)\ \right)\ \ J(b, a) =  \ J_{\nu'}(b) J_{\nu}^2(a)\ \ .
\end{equation}
Given equation (\ref{eq=gaugexcrit}) for $J_{\nu'}(x_{am\nu'})$ and equation  
(\ref{eq=Iode}) for the normalization integral $I_\nu(a)$, we find the solution
\begin{equation}
J(x_{am\nu'}, a ) = I_\nu(a) J_{\nu'}(x_{am\nu'})
\end{equation}
so that
\begin{equation}
{\cal M}_{nmn} =  \frac{2}{k} \  {\rm a}_m {\rm h}_n^2\ e^{3kR\pi}\ I_\nu(x_{n\nu}) J_{\nu'}(x_{am\nu'})\ \ ,
\end{equation}
which gives, from equations (\ref{eq=norms}) and (\ref{eq=specI}),
\begin{equation}
{\cal M}_{nmn} =  k^2 \left(\ 1 - \frac{m_{a,5}^2}{k^2x_{am\nu'}^2}\ \right)^{-1/2} \ \ .
\end{equation}
Thus the effective 4D charge coupling of a Fermi Kaluza-Klein mode
$\psi_n$ to a massive gauge mode $A_{am}$ is
\begin{equation}
e_{am,{\rm 4D}} = e_{a5}\,{\cal M}_{nmn} = e_{a5}\, k^2 \left(\ 1 - \frac{m_{a,5}^2}{k^2x_{am\nu'}^2}\ \right)^{-1/2}\ . \end{equation}

Note that this is universal: all Fermi Kaluza-Klein modes $\psi_n$
couple to the massive gauge species $A_{am}$ with a charge independent
of the fermion species, its 4D mass $m_n$, or its bulk mass $m_5$. The
4D charge coupling does depend on gauge boson mass however: both
through its bulk mass $m_{a,5}$, and through its particular 4D
Kaluza-Klein mass $m_m$, through $x_{am\nu'}$. Thus fermions couple to
$W^{\pm}$ and $Z$ bosons with mildly different charges, and to the
higher Kaluza-Klein $W^{\pm}$ and $Z$ modes with different charges still,
approaching the single limit $e_{a,{\rm 4D}} = e_{a5}\, k^2$ for all
high-lying gauge Kaluza-Klein modes.

\section{Conclusions\label{conclude}}

We treat the Randall-Sundrum solution as a brane-warped gravitational
background for a fully extradimensional field theory of gravity and
matter. This enables complete and self-consistent tracking of the 4D
effective physics, both particle and gravitational, induced by brane
formation. Our theory's dimensional reduction in the warped
Randall-Sundrum background is remarkably successful. Hierarchy
solution, fully calculable, remains robust. Moreover, the 4D effective
field theory induced for matter is a unique exemplar of a
dimensionally reduced, integrated effective field theory that preserves 4D
charge universality. This 4D effective theory is unusual, in
describing physics perceived due to a small, unresolved warped extra
dimension. Nonlocalization, or localization to different branes, of
differing 5D fields remains irrelevant below near-Planckian energies;
the unresolvable warped extra dimension simply induces the single
integrated 4D effective field theory presented here. Specifically,
this theory has Planck scale gravity, ${\rm TeV}$ scale matter,
universal charges, and an unresolved extra dimension of size $12\, M_{\rm
Pl}^{-1}$.

In this theory, the smallness of the warped extra dimension is
essential to solving the hierarchy problem. This required smallness
enables the viability of an integrated 4D effective theory, even
without localizing all standard matter fields to a single brane. Other
warped extradimensional models, in other brane-defect backgrounds and
dimensionalities, could lead instead to large warped extra dimensions,
where the relevant induced 4D effective field theories must be those
confined to a single brane. We examine such possibilities --- warped
extradimensional models, and the 4D effective theories they induce for
both matter and gravity --- generically in \cite{warpgen}. The
accomplishment here though, in the Randall-Sundrum background, remains
substantial.  String theorists require a pile-up of multiple
collocated branes to establish nonabelian gauge theories on the brane;
requiring 2 branes with small separation, then, seems a small price
for field theorists to pay, to attain a fully field theoretic model of
extra dimension, whose dimensional reduction demonstrably solves both
hierarchy and charge universality problems.

\acknowledgments We are grateful to Eduardo Ponton, Marcela Carena, and Sekhar
Chivukula for helpful discussions.  We thank the Radcliffe Institute
for Advanced Study for support, and the Fermilab astro-particle theory
group for hospitality, during portions of this work.


\begin{thebibliography}{9999}
\bibitem{fn1} Technically, in \cite{rs1}, Randall-Sundrum offer two
variant views of this fundamental scale. It can be viewed as {\rm TeV}
scale, as I've defined here, or it can be viewed as Planck scale, via
field redefinition. This subtlety does not impact the argument here,
that Randall-Sundrum give an unconventional spatial interpretation of
a 4D effective theory, as universal and omnipresent in the extra
dimension yet perceived pointwise, relative to a distinguishable
local metric. In viewing the Planck scale instead  as the simultaneous
fundamental scale, the particle 4D effective theory assumes this
difficulty, instead of 4D effective gravity.
\end{thebibliography}
\end{document}